\documentclass{webofc}
\usepackage[varg]{txfonts}
\usepackage{mathtext}
\usepackage{amsfonts}
\usepackage{amsmath}
\usepackage{amssymb}
\usepackage{amscd}
\usepackage{cancel}
\usepackage{graphicx}
\usepackage[english]{babel}

\begin{document}
\title{\vspace{0cm}Towards wormhole beyond Horndeski}

\author{\firstname{S.} \lastname{Mironov}\inst{1,2}\fnsep\thanks{\email{sa.mironov_1@physics.msu.ru}} \and
        \firstname{V.} \lastname{Rubakov}\inst{1,3}\fnsep\thanks{\email{rubakov@inr.ac.ru}} \and
        \firstname{V.} \lastname{Volkova}\inst{1,3}\fnsep\thanks{\email{volkova.viktoriya@physics.msu.ru}}
}

\institute{Institute for Nuclear Research of the Russian Academy of Sciences,
60th October Anniversary Prospect, 7a, 117312 Moscow, Russia
\and
           Institute for Theoretical and Experimental Physics,
Bolshaya Cheriomyshkinskaya, 25, 117218 Moscow, Russia
\and
           Department of Particle Physics and Cosmology, Physics Faculty,
M.V. Lomonosov Moscow State University,
Vorobjevy Gory, 119991 Moscow, Russia
          }

\abstract{
  We address the issue of whether a no-go theorem for static,
  spherically symmetric wormholes,
proven in Horndeski theories, can be circumvented
by going beyond Horndeski. We show that the ghost 
instabilities
which are at the heart of the no-go theorem, can indeed be avoided.
The wormhole solutions with the latter property are, however,
strongly fine tuned, and hence it is likely that they are
unstable. Furthermore, it remains unclear
whether these solutions have other pathologies, like
gradient instabilities along angular and radial directions.
}

\maketitle

\section{Introduction and summary}
Horndeski theories \cite{Horndeski} can
violate the Null Energy Condition (NEC) in a healthy way.
Their distinctive feature is that despite the presence of second derivatives in the Lagrangian, equations of motion are
second order. Similar properties, but  at the level of 
unconstrained perturbations,
hold in beyond Horndeski theory \cite{Gleyzes:2014dya}. Because of
potentially healthy NEC violation,
Horndeski and beyond Horndeski theories
have been recently explored
for constructing various cosmological solutions,
for instance, bounce and Genesis.
However, there are no-go
theorems \cite{1605.05992,1606.05831,1607.04099,1701.02926}
stating that classically stable, at all times,
bouncing or Genesis
solutions are absent in
Horndeski theory. On the other hand,
these theorems
can be circumvented  in beyond Horndeski theory,
and a fully stable bouncing solution can be
constructed \cite{1610.03400,1610.04207,1705.03401,1705.06626}.

Similar issues arise in the static, spherically symmetric context.
The analog of the cosmological bounce in this case is
traversable wormhole.
One might expect, by analogy to cosmological setting,
that there is a no-go theorem in Horndeski theory.
This expectation is confirmed in Horndeski theory: static, spherically
symmetric, asymptotically flat wormholes are plagued with ghost
and/or radial gradient instability \cite{1601.06566,1711.04152}.
By the same analogy, one might expect that the no-go theorem
might be circumvented in beyond Horndeski theories, so that a fully stable
wormhole might exist. On the other hand, censorship
against the time machine~\cite{Morris:1988cz,Morris:1988tu}
is also a weighty argument.

In this paper we address the latter issue, and concentrate on
ghost instabilities.
We find that, indeed, there exist wormholes
without these instabilities in beyond Horndeski theories.
However, these solutions are
strongly fine tuned. Our current understanding is that
this fine tuning actually implies wormhole instability at
non-linear level. Furthermore, we have not yet studied in full
detail possible gradient and tachyon instabilities; although
off hand there appears no reason for the existence of the
latter, there might be surprizes here as well.

\section{NEC, bouncing Universe and wormhole}
 The
NEC is a fairly general property of matter degrees of freedom, which reads
\begin{equation}
\label{ur01}
T_{\mu\nu}k^{\mu}k^{\nu}\geq0,
\end{equation}
where $k^{\mu}$ is an arbitrary null vector and $T_{\mu\nu}$ is energy-momentum
tensor. The assumption that the NEC holds is crucial for
Penrose theorem \cite{Penrose} that states that
the presence of a trapped surface unavoidably leads to a singularity.
In the cosmological context
this means that contracting Universe ends up in a singularity
and, by time reversal, that expanding Universe has
a singularity in the past. Likewise, Penrose theorem
and its generalizations
forbid
the existence of Lorentzian wormholes~\cite{Hochberg:1998ha}.

It is well-known that healthy NEC violation
requires very peculiar types of matter, see \cite{rub} for a review.
For a long time there was even a belief that breaking NEC always
leads to appearance of pathological degrees of freedom
(under pathological degrees of freedom one usually means
gradient instabilities and/or ghosts). The situation has changed
fairly recently, when Horndeski theories (generalized Galileons)
became a subject of intense
study~\cite{Luty:2003vm,Nicolis:2004qq,Nicolis:2008in,Genesis1}.

The (imperfect) analogy between bouncing Universe and static, spherically
symmetric wormhole becomes clear once one writes (spatially flat)
cosmological FLRW metric:
\begin{equation}
  d s^2 =N^2 dt^2-
  a(t)^2\left(dr^2+r^2\left(d\theta^2+\sin^2\theta\,d\varphi^2\right)\right)
  \, ,
\end{equation}
and static spherically symmetric metric:
\begin{equation}
  d s^2=A(r)dt^2-\frac{dr^2}{B(r)}-R(r)^2
  \left(d\theta^2+\sin^2\theta\,d\varphi^2\right) \, .
\label{sep2-18-1}
\end{equation}
Modulo signs here and there, the only
difference is that functions $A,~B,~R$ depend on radius but not time.
The profiles
of $a(t)$ and $R(r)$ for the bouncing Universe and wormhole,
respectively, are quite similar (where on the right panel
we think of $R\rightarrow+\infty$ as our Universe
and $R\rightarrow-\infty$ as another Universe or very distant part of ours):
$$
\includegraphics[height=4.5cm]{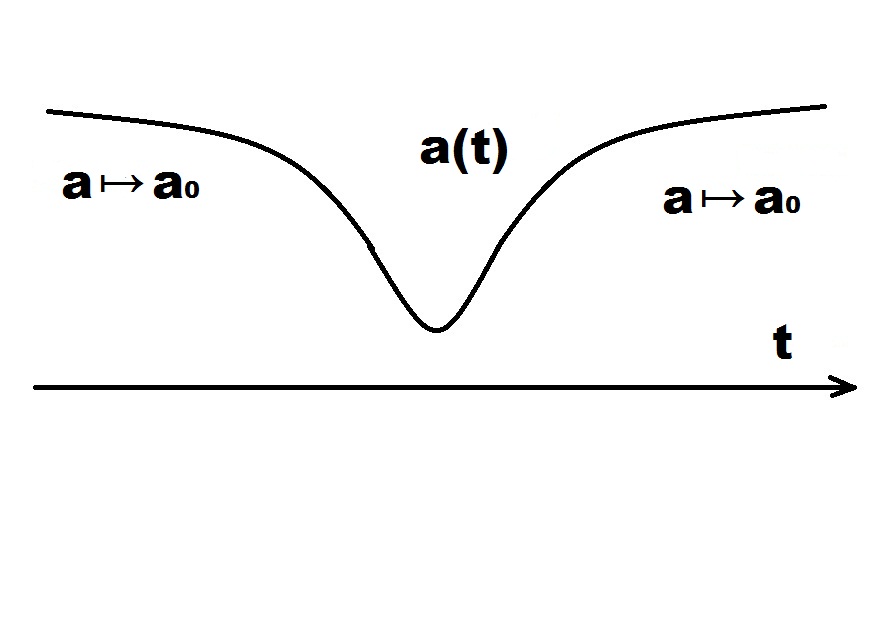}
\includegraphics[height=4.5cm]{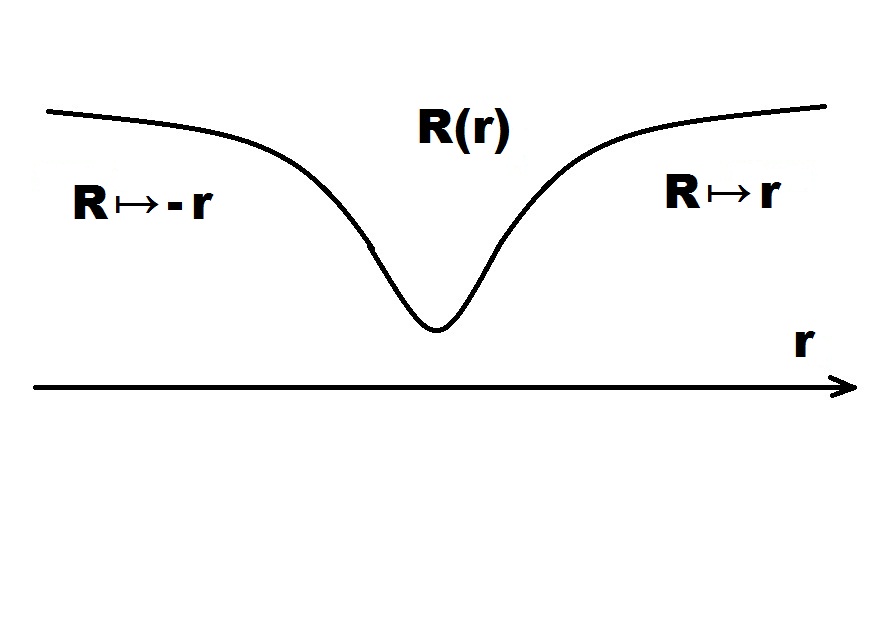}
$$
So, it is not surprising that theories that do not admit fully stable
bouncing
cosmologies, do not support stable static spherically symmetric wormholes
as well, and vice versa.
Gradient instabilities in bouncing
cosmologies (wrong sign of spatial gradient terms in the quadratic
action for small perturbations) have their counterparts as ghost
instabilities of wormhole backgrounds (wrong sign of terms with
time derivatives). This is precisely what happens in Horndeski
theories  (cf. Refs.~\cite{1605.05992,1606.05831}
for bouncing Universe and Refs.~\cite{1601.06566,1711.04152} for wormholes).
An intriguing question is whether the opposite is correct as well:
beyond Horndeski theories do have completely stable bouncing
solutions~\cite{1610.03400,1610.04207,1705.03401,1705.06626}, do they
support wormholes?

\section{Horndeski and beyond}

Horndeski theory is the most general scalar-tensor theory
of gravity whose Lagrangian contains  second derivatives, and yet
the field equations are second order in derivatives.
If one allows third derivatives in the equations of motion,
but restricts the unconstrained equations for perturbations to have only
second derivatives, then in the simplest cases
one comes to beyond Horndeski Lagrangians.
The most general forms of these Lagrangians are
 \begin{equation} \begin{array}{l}
    S=\int\mathrm{d}^4x\sqrt{-g}\left(\mathcal{L}_2 + \mathcal{L}_3 + \mathcal{L}_4 + \mathcal{L}_5 + {\bf \mathcal{L_{BH}}}\right), \\
    \\
    \mathcal{L}_2=F(\pi,X), \\
    \\
    \mathcal{L}_3=K(\pi,X)\Box\pi, \\
    \\
    \mathcal{L}_4=-G_4(\pi,X)R+2G_{4X}(\pi,X)\left[\left(\Box\pi\right)^2-\pi_{;\mu\nu}\pi^{;\mu\nu}\right], \\
    \\
    \mathcal{L}_5=G_5(\pi,X)G^{\mu\nu}\pi_{;\mu\nu}+\frac{1}{3}G_{5X}\left[\left(\Box\pi\right)^3-3\Box\pi\pi_{;\mu\nu}\pi^{;\mu\nu}+2\pi_{;\mu\nu}\pi^{;\mu\rho}\pi_{;\rho}^{\;\;\nu}\right], \\
    \\
    {\bf \mathcal{L_{BH}}}={\bf F_4}(\pi,X)\epsilon^{\mu\nu\rho}_{\quad\;\sigma}\epsilon^{\mu'\nu'\rho'\sigma}\pi_{,\mu}\pi_{,\mu'}\pi_{;\nu\nu'}\pi_{;\rho\rho'} +{\bf F_5}(\pi,X)\epsilon^{\mu\nu\rho\sigma}\epsilon^{\mu'\nu'\rho'\sigma'}\pi_{,\mu}\pi_{,\mu'}\pi_{;\nu\nu'}\pi_{;\rho\rho'}\pi_{;\sigma\sigma'},
    \\
  \end{array}
  \end{equation}
 where $\pi$ is the scalar (Galileon) field,
 $X=g^{\mu\nu}\pi_{,\mu}\pi_{,\nu}$,
 $\pi_{,\mu}=\partial_\mu\pi$,
 $\pi_{;\mu\nu}=\triangledown_\nu\triangledown_\mu\pi$,
 $\Box\pi = g^{\mu\nu}\triangledown_\nu\triangledown_\mu\pi$,
 $G_{4X}=\partial G_4/\partial X$, etc. The Lagrangians
 $\mathcal{L}_2$~--~$\mathcal{L}_5$ give Horndeski theory,
 while  ${\bf \mathcal{L_{BH}}}$  is characteristic of beyond Horndeski.

\section{No-go theorem}
In this section we briefly consider no-go theorem
that forbids stable wormhole solutions in Horndeski theory~\cite{1711.04152}.
Very similar theorem holds for spatially flat bouncing
Universe in this theory~\cite{1605.05992,1606.05831}.
We give the following lemma first.
Suppose we have a "nice" function
$f(x)$ defined for all $x$ from $-\infty$ to $\infty$.
Now, if $f'(x)>\epsilon>0$, then
$f(x_0)=0$ at some point $x_0$.
Let us illustrate this in the following plots:
\begin{center} $\includegraphics[height=2.9cm]{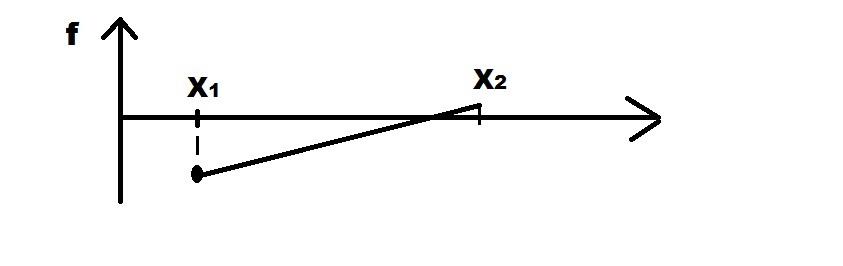}$

  If $f(x_1)<0\, ,~~~~$ then $f(x_2)\geq 0$,
  where $x_2=x_1-\frac{f(x_1)}{\epsilon}$.

 $\includegraphics[height=2.9cm]{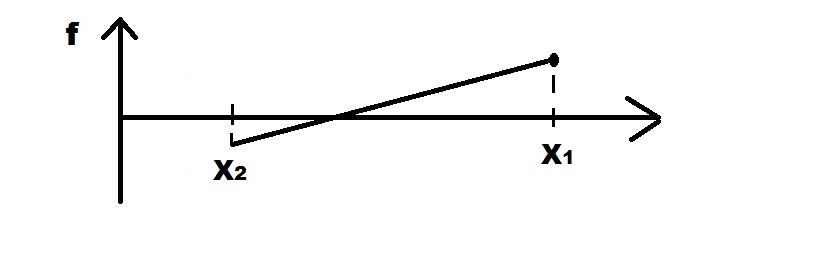}$

 Alternatively, if $f(x_1)>0\, ,~~~~$ then $f(x_2)\leq 0$, where $x_2=x_1-\frac{f(x_1)}{\epsilon}$.
\end{center}
In fact, the function $f(x)$ may be singular at one or more points, as
shown below:
\begin{center}
 $\includegraphics[height=4.5cm]{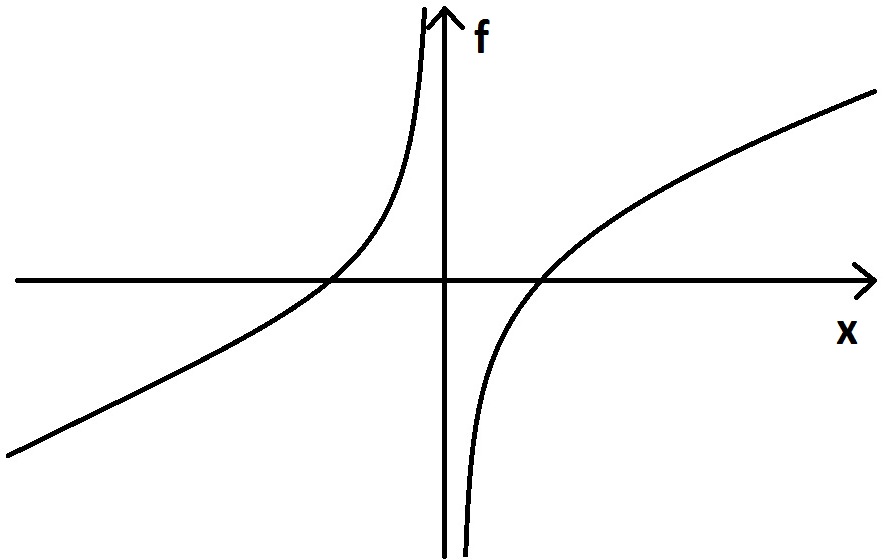}$
\end{center}

Having this in mind, we consider perturbations of metric and scalar field
in the spherically symmetric background \eqref{sep2-18-1}. We begin with
Horndeski theory, ${\bf F_4}(\pi,X) = {\bf F_5}(\pi,X)=0$. 
Quadratic Lagrangian for perturbations has been derived by
Kobayashi, Motohashi and
Suyama~\cite{Kob1,Kob2} who make use of Regge--Wheeler approach.
In the spherically symmetric background,
perturbations are classified into parity even and parity odd sectors.
There are
two fields $v^i$ ($i=1,2$) in parity even sector, and their Lagrangian
is
\begin{equation}
  {\cal L}_{even}=\frac{1}{2} {\cal K}_{ij} {\dot v}^i {\dot v}^j-\frac{1}{2}
  {\cal P}_{ij} {v^i}' {v^j}'-{\cal Q}_{ij} v^i {v^j}'-\frac{1}{2}
  {\cal M}_{ij} v^i v^j,
  \label{sep2-18-2}
\end{equation}
where the coefficients  ${\cal K}_{ij} (r)$,  ${\cal P}_{ij}(r)$,
${\cal Q}_{ij}(r)$,  ${\cal M}_{ij}(r)$ are expressed through the functions
in the Horndeski Lagrangian evaluated on the spherically symmetric background,
and prime denotes $\partial/\partial r$. Expression \eqref{sep2-18-2}
is written for spherical harmonics $v^i_{lm}$, and the functions
 ${\cal K}_{ij} (r)$,  etc., depend also on angular momentum $l$.

As we already pointed out, we concentrate on ghost
instability issue.
The absence of this instability in parity even sector requires
\begin{equation}
\label{stab_even}
  {\cal K}_{11}>0,\quad\det ({\cal K}) > 0,
\end{equation}
(the absence of gradient instabilities,
which we do not consider for the even parity modes,
requires  similar
relations for ${\cal P}_{ij}(r)$,
${\cal Q}_{ij}(r)$ and  ${\cal M}_{ij}(r)$).

For the odd parity modes the quadratic Lagrangian reads
\begin{equation}
\label{lagr_odd}
{\cal L}_{odd} = \frac{l(l+1)}{2(l-1)(l+2)}\sqrt{\frac{B}{A}} R^2 \left[ \frac{\mathcal{H}^2}{\sqrt{A}\mathcal{G}} \dot{Q}^2 - \frac{\sqrt{B}\mathcal{H}^2}{\mathcal{F}} (Q')^2 -\frac{l(l+1)}{R^2}\cdot \mathcal{H} Q^2 - V(r) Q^2 \right],
\end{equation}
so we have similar restrictions, the
relevant ones for the no-go theorem being
\begin{equation}
\label{stab_odd}
\mathcal{F}>0, \quad \mathcal{H}>0 ,
\end{equation}
where, again, $\mathcal{F} (r)$ and $\mathcal{H} (r)$ are
expressed through the Lagrangian functions
evaluated on the spherically symmetric background
Also, background metric coefficients cannot cross zero:
\begin{center}$A>0$, $B>0$, $R>0$. \end{center}
Now, the key point is that modulo a manifestly positive factor,
$\det K$ reads
\begin{equation}
\det K \sim \mathcal{F}(2\xi'-\mathcal{F})>0 \; ,
\label{sep2-18-3}
\end{equation}
where $\xi(r)$ has the following structure
\begin{equation}
\label{ur1}
\xi=\left[\frac{(R\mathcal{H})^2}{\Theta}\right],
\end{equation}
with
\begin{align*}
  \mathcal{H} &=2 G_4 +2 G_{4X} B \pi'^2,
\\
  \Theta &= 2 \mathcal{H} R R' + \Xi \pi',
 \end{align*}
 where
 \begin{equation*}
   \Xi = -2 G_{4\pi} R^2-4 G_{4X} B R R' \pi'+2 G_{4\pi X} B R^2 (\pi')^2+4 G_{4XX} B^2 R R' (\pi')^3,
\end{equation*}
and we set $G_5(\pi,X)=0$ for the sake of brevity.   
Equations (\ref{sep2-18-3}), (\ref{ur1}) enable one
to prove the no-go theorem \cite{1607.04099,1601.06566,1711.04152}.
Indeed, in
full analogy with the homogeneous case in Refs.~\cite{1606.05831},
$\xi$ cannot cross zero, but it has to,
since its derivative is always positive (see the lemma above with $\xi$ being a counterpart of function $f(x)$).

\section{What becomes of no-go theorem for wormholes beyond Horndeski}
The no-go theorem does not work
for wormholes beyond Horndeski theory.
In spherically symmetric background
(\ref{sep2-18-1}),
like in the cosmological case~\cite{1610.04207,1705.03401,1705.06626},  both relations   (\ref{sep2-18-3}) and
(\ref{ur1}) get modified. Namely, the analog of (\ref{ur1}) reads
\begin{equation}
\xi=\left[\frac{R^2\mathcal{H}(\mathcal{H}-\mathcal{D})}{\Theta}\right],
\end{equation}
where the expressions for   $\mathcal{H}$ and   $\Theta$  are now (we set $G_5={\bf F_5}=0$ in what follows)
\begin{align*}
  \mathcal{H} & = 2 G_4 +2 G_{4X} B \pi'^2 + 2{\bf F_4} B^2 \pi'^4,\\
  \Theta &= 2 \mathcal{H} R R' + \Xi \pi',
  \end{align*}
  with
  \begin{align*}
  \Xi = &-2 G_{4\pi} R^2-4 G_{4X} B\: R R' \pi'+2 G_{4\pi X} B R^2 (\pi')^2+4 G_{4XX} B^2\: R R' (\pi')^3 \\ 
  &-16 {\bf F_4} B^2\: R R' (\pi')^3+4 {\bf F_{4X}} B^3\: R R' (\pi')^5, 
  \end{align*}
  while
\begin{equation}
  \mathcal{D} = 2 {\bf F_4} B^2 \pi'^4.
\end{equation}
The requirement that $\det K > 0$ is now 
\begin{equation}
\label{new_det}
  \det K \sim (\mathcal{F}-\mathcal{C}\mathcal{Q})(2\xi'-\mathcal{F})-
        \mathcal{C}\mathcal{Q}^2 >0,
\end{equation}
where
\begin{equation}
\mathcal{Q} = a_1 \cdot \frac{\left( a_2 \mathcal{D} \right)'}{R'}, \quad \mathcal{C} = \frac{4l(l+1)}{(l+2)(l-1)},
\end{equation}
and both $a_1$ and $a_2$ are manifestly positive: $a_2 = a_1^{-1} \cdot R/2= \sqrt{A}/(\pi' \sqrt{B})$.
Assuming that asymptotically, as $r\to \pm \infty$, the theory reduces to GR,
we have $ \mathcal{Q} \to 0$, so that $\mathcal{F}-\mathcal{C}\mathcal{Q} > 0$
 as $r\to \pm \infty$.
Positivity of $\det K$ requires that  $\mathcal{F}-\mathcal{C}\mathcal{Q}$ does not
change sign, so we still have to require that $\xi^\prime > \frac12 \mathcal{F} > 0$ at all $r$.
However,
again in full analogy with the cosmological 
case~\cite{1606.05831,1705.06626},
$\xi$ can safely cross zero in beyond Horndeski theory
due to the additional term $\mathcal{D}$, so the no-go theorem
no longer works.

An explicit example of such a solution is shown in
Fig.~$1$ (for the sake of brevity
we do not present explicitly all Lagrangian functions).
For the wormhole construction we make use of functions $F$, $G_{4}$ and ${\bf F_{4}}$ only, and set $K = G_5 = {\bf F_{5}} = 0$. We choose 
$$
R=\sqrt{1+r^2}, \quad A=B=1,
$$
in metric~\eqref{sep2-18-1} and
$$
\Theta = r/3,\quad \mathcal{H}=1,\quad \mathcal{F}=1,\quad \mathcal{D}=\mbox{cosh}^{-2}(r).
$$ 
Then we find the Lagrangian functions $F$, $G_4$, ${\bf F_{4}}$ by requiring that the field equations are satisfied. These Lagrangian functions are smooth and we have the following asymptotics as
$r \rightarrow \pm \infty$:
\begin{equation}
 G_{4}(\pi,X) \rightarrow 1/2.
\end{equation} 
Hence, gravity asymptotically reduces to General Relativity. Our Lagrangian functions ensure that $\mathcal{F}$, $\mathcal{H}$, 
$\mathcal{G}$, ${\cal K}_{11}$ and
$\det ({\cal K})$ are strictly positive. Thus, the solution appears to be ghost-free in the parity even sector and free of both gradient and ghost instabilities in the parity odd sector. We note that our solution as it stands is not guaranteed to be stable against all linear perturbations. Namely, we have not studied gradient instabilities in the parity even sector.

The worst drawback of this solution is, however, strong fine tuning. One has to ensure that $\left( a_2 \mathcal{D}\right)'=0$ at exactly the same point where $R'=0$ (the wormhole throat), otherwise $\mathcal{Q}$ is singular and the constraint $\det K>0$ is not satisfied, see eq.~\eqref{new_det}. 
Thus, the solution reveals its unhealthy nature: at least naively, the attempts to introduce asymmetric deviations into the solution result in the development of ghost instabilities.
\begin{figure}[h!]
\label{f1}
\begin{center}
\includegraphics[width=0.6\linewidth]{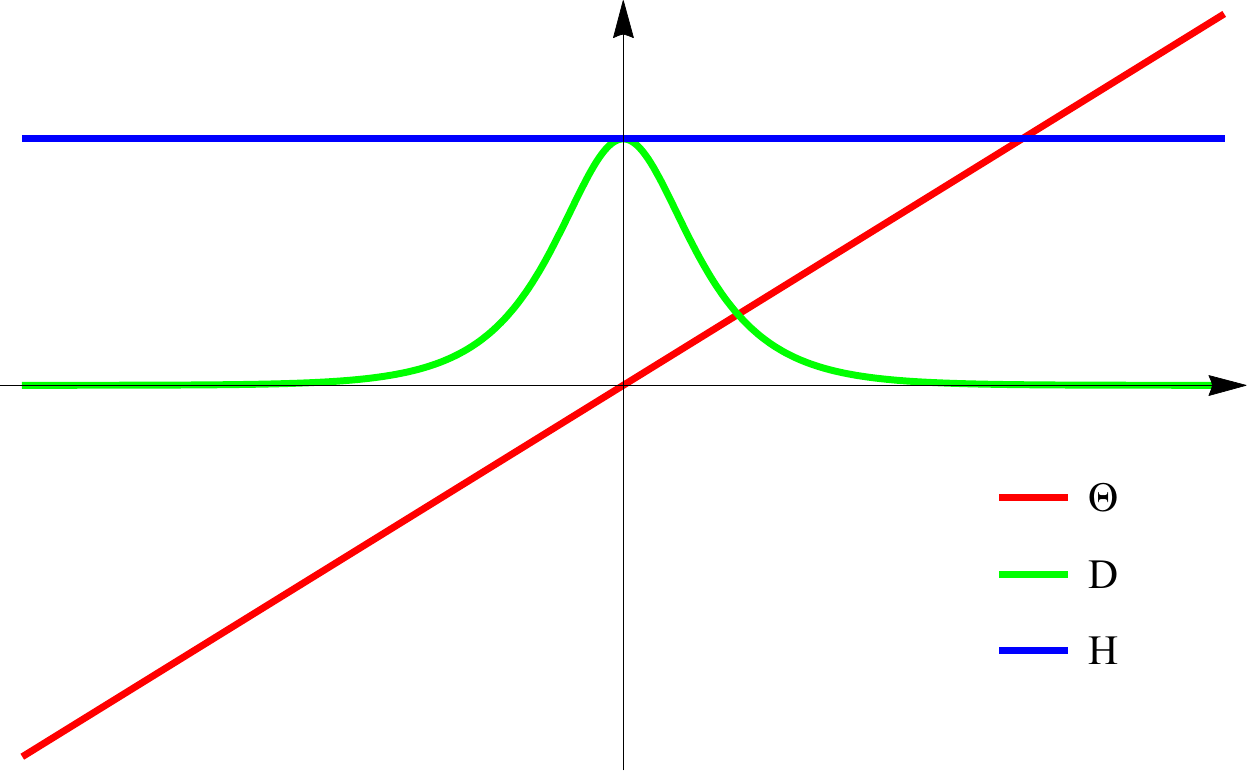}
\caption{Wormhole solution: $\Theta$ is a linear function of radius, $\mathcal{H}$ is a constant and $\mathcal{D} = \cosh^{-2}(r)$.}
\end{center}
\end{figure}

\section{Acknowledgements}
This work has been supported by Russian Science Foundation grant 14-22-00161.

\end{document}